\documentclass[prb,twocolumn,superscriptaddress]{revtex4-1}
\usepackage{amsmath,amssymb,graphicx,color,bm,epstopdf,float,caption,subcaption,tabularx,multirow}
\usepackage{scrextend}
\usepackage{hyperref}
\renewcommand{\vec}[1]{\mbox{\mathversion{bold}$#1$}}

\begin{document}

\title{
Electro-absorption of silicene and bilayer graphene quantum dots 
}
\author{Hazem Abdelsalam}
\email{hazem.abdelsalam@etu.u-picardie.fr}
\affiliation{University of Picardie, Laboratory of Condensed Matter
Physics, Amiens, 80039, France}
\affiliation{Department of Theoretical Physics, National Research Center, Cairo,12622, Egypt}

\author{Mohamed H. Talaat}
\affiliation{Physics Department, Faculty of Science, Ain Shams University, Cairo, Egypt}

\author{Igor Lukyanchuk}
\affiliation{University of Picardie, Laboratory of Condensed Matter
Physics, Amiens, 80039, France}
\affiliation{L. D. Landau Institute for Theoretical Physics, Moscow, Russia}

\author{M. E. Portnoi}
\affiliation{School of Physics, University of Exeter, Stocker Road, Exeter EX4 4QL, United Kingdom}

\author{V. A. Saroka}
\email{v.saroka@exeter.ac.uk}
\affiliation{School of Physics, University of Exeter, Stocker Road, Exeter EX4 4QL, United Kingdom}
\affiliation{Institute for Nuclear Problems, Belarusian State University, Bobruiskaya 11, 220030 Minsk, Belarus}

\date{\today}
\begin{abstract}

We study numerically the optical properties of low-buckled silicene and AB-stacked bilayer graphene quantum dots subjected to an external electric field, which is normal to their surface. Within the tight-binding model, the optical absorption is calculated for quantum dots, of  triangular and hexagonal shapes, with zigzag and armchair edge terminations.
We show that in triangular silicene clusters with zigzag edges a rich and widely tunable infrared absorption peak structure originates from transitions involving zero energy states. The edge of absorption in silicene quantum dots undergoes red shift in the external electric field for triangular clusters, whereas blue shift takes place for hexagonal ones. In small clusters of  bilayer graphene with zigzag edges the edge of absorption  undergoes blue/red shift for triangular/hexagonal geometry. In armchair clusters of silicene blue shift of the absorption edge takes place for both cluster shapes, while red shift is inherent for both shapes of the bilayer graphene quantum dots.


\end{abstract}
\maketitle

\section{Introduction}

Non-planar graphene-derivative materials have attracted considerable
attention \cite{Ni2014,Tsai2013,Kumar2011,Liang2012,Artyukhov2010,Lui2011,Lukyanchuk,Ezawa2015} because of
their tunable electronic properties, different from those of the single-layer
graphene. Application of the electric field, $E$, across the bilayer (multilayer)
graphene system opens a gap between the conduction and valence bands. 
\cite{McCann,ECastro2007,Ohta2006,Khodkov2015} The same also happens with
silicene because of the buckling of its honeycomb lattice.\cite{Tsai2013,Kumar2011,Liang2012,Lui2011} The atoms of the type A and B of the
lattice are displaced alternatively in the vertical direction and are
subjected to a different, electric field producing, potential gradient. The
possibility of controlling the gap offers a wealth of new routes for the
next generation of field effect transistors and optoelectronic devices.\cite{Ni2014,Tsai2013,koshino2013} However, the on-chip nano-scale realization
of such devices requires finite-size components like nanoribbons and nanoflakes or quantum dots (QDs).\cite{Kvashnin2015} Therefore, a deeper understanding of their individual electronic properties, which can be
substantially different from those in infinite systems because of the
finite-size electronic confinement,\cite{Sorokin2008} is needed.


The electronic properties of various graphene nanoribbon structures and the influence of the applied voltage is being studied both for the out-of-plane\cite{Sahu2008,Sahu2010,Yu2013} and for the in-plane \cite{Huang2008,Chang2006,Saroka2014,Saroka2015} field directions. The optical and magnetic
properties of the single and multilayer graphene QDs of various shapes have
also been studied at zero field.\cite{Ezawa,Zhang2008,Chernozatonskii2013,Rossier2007,Potasz,Kosimov,Espinosa,Espinosa2,Hazem,Guclu2016,Zarenia2011,Costa2014}
The distinctive property of these QDs is the opening of a
finite-size energy gap due to the electron confinement, that is different
from the above mentioned field-induced gap since it exists also at $E=0$. In
addition, the novel electronic states localized at the sample boundary are
formed.\cite{Ezawa,Zhang2008,Rossier2007} In the energy spectrum these
states are located inside the gap in the vicinity of the zero energy. This corresponds to the Dirac point, when size of the system tends to infinity,
therefore they are usually referred to as zero energy states (ZES). 
Unlike the ZES in single layer graphene QDs, the ZES in silicene and bilayer
graphene QDs can be easily manipulated by an electric field applied
normally to the graphene or silicene layers.\cite{Guclu2011,Hazem2,Costa2015,Costa2016} 

In this paper we explore this functionality for the design of the QD-based
optoelectronic devices. We discuss the effect of an electric
field on the optical absorption cross section in silicene and bilayer
graphene QDs and how the applied field can control the number and
intensities of absorption peaks. 

 This paper is organized as follows: in Section~\ref{Sec:Model} we introduce structure classification and provide details of our tight-binding calculations. In Section~\ref{Sec:Res} we present and discuss optical absorption spectra in electric field for a range of QD types. Finally, our discussion is summarised in Section~\ref{Sec:Conc}.

\section{\label{Sec:Model}Structures and calculation model}

\begin{figure}[tbp]
	\centering
\begin{subfigure}{.23\textwidth}
  \centering
  \includegraphics[width=\textwidth]{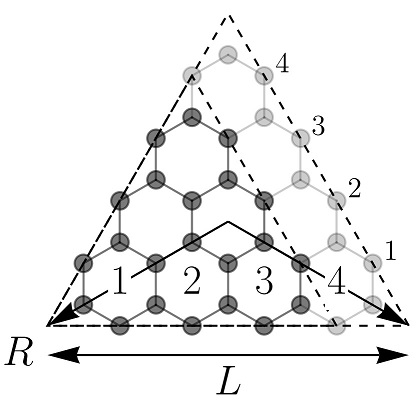}
  \caption{}
 \end{subfigure}%
\hfill
\begin{subfigure}{.23\textwidth}
  \centering
  \includegraphics[width=\textwidth]{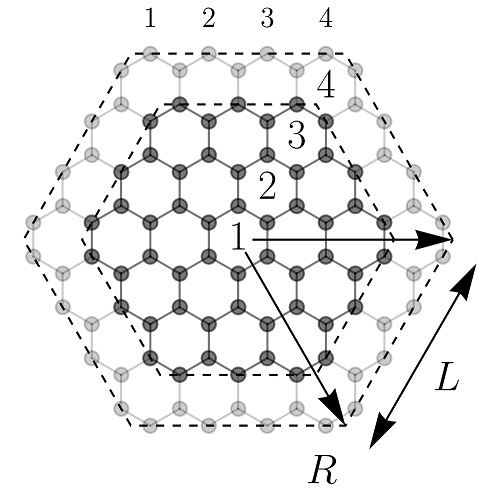}
  \caption{}
\end{subfigure}

\begin{subfigure}{.23\textwidth}
  \centering
  \includegraphics[width=\textwidth]{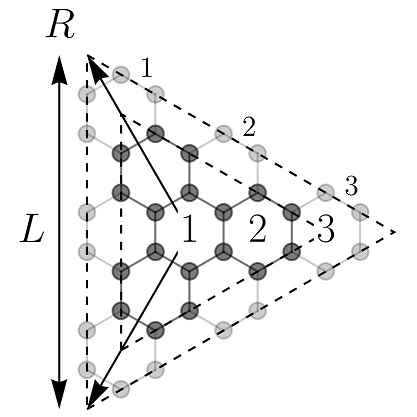}
  \caption{}
 \end{subfigure}%
 \hfill
\begin{subfigure}{.23\textwidth}
  \centering
  \includegraphics[width=\textwidth]{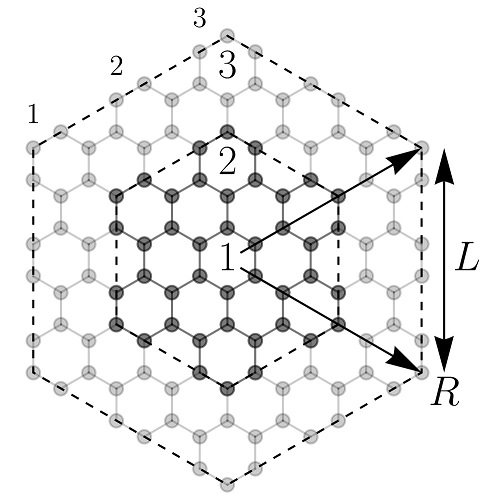}
  \caption{}
\end{subfigure}
\caption{The four main types of QD, based on the 2D hexagonal lattice: (a) zigzag triangular, (b) zigzag hexagonal, (c) armchair triangular , (d) armchair hexagonal, where  $R$ and $L$ are  the circumscribed circle radius and edge length, respectively. Quantum dot indexing is presented by larger and smaller font numbering.}
\label{fig:Classification}
\end{figure}
\begin{figure}[tbp]
	\centering
\begin{subfigure}{.25\textwidth}
  \centering
  \includegraphics[width=\textwidth]{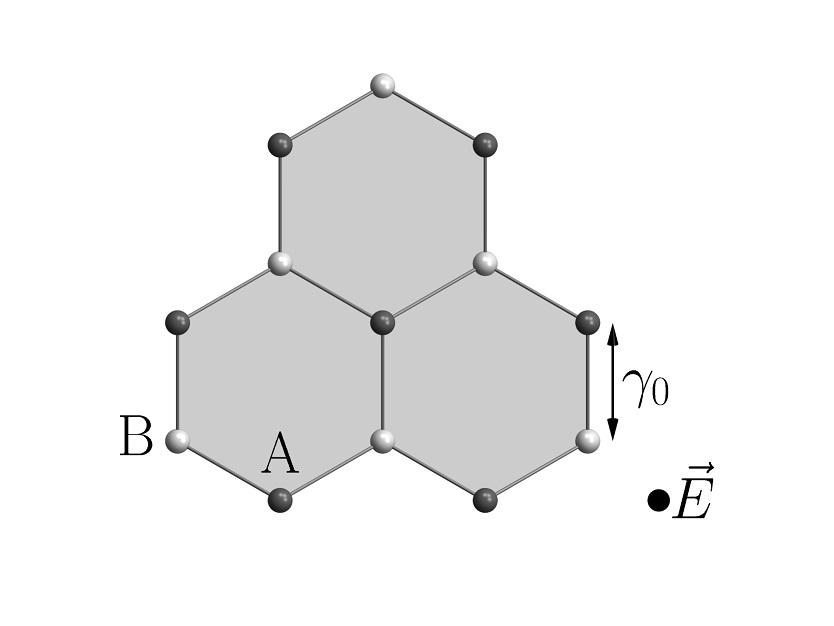}
  \caption{Top view}
\end{subfigure}%
\begin{subfigure}{.25\textwidth}
  \centering
  \includegraphics[width=\textwidth]{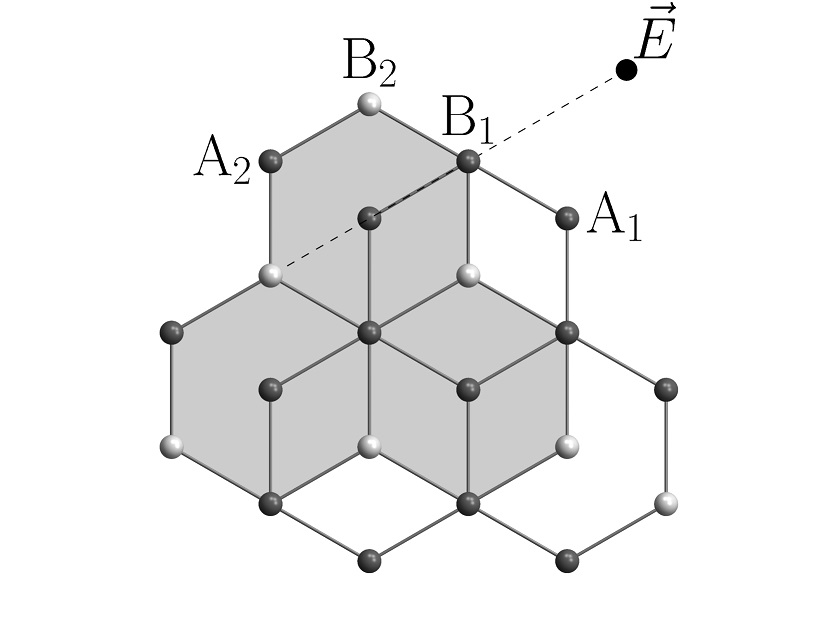}
  \caption{Top view}
\end{subfigure}%

\begin{subfigure}{.25\textwidth}
  \centering
  \includegraphics[width=\textwidth]{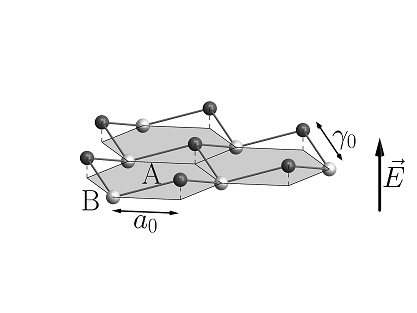}
  \caption{Side view}
\end{subfigure}%
\begin{subfigure}{.25\textwidth}
  \centering
  \includegraphics[width=\textwidth]{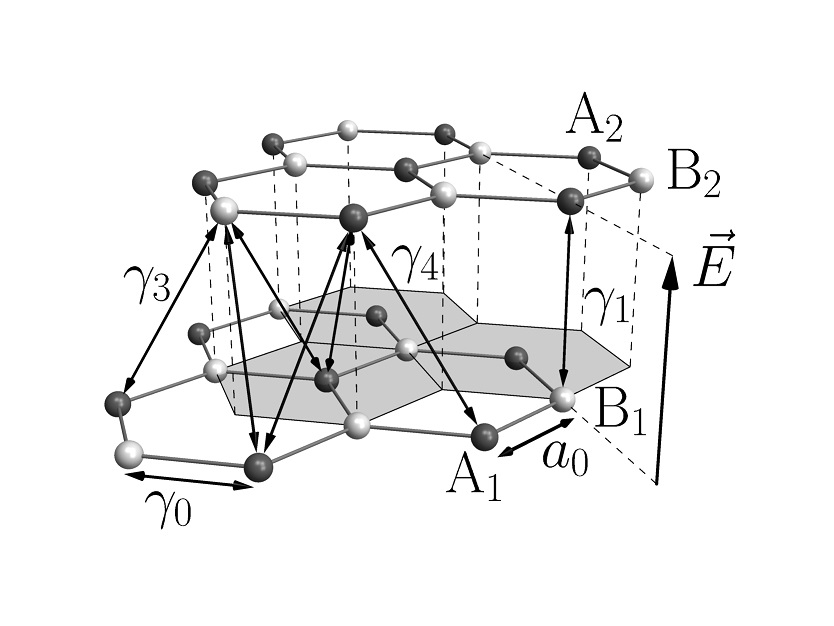}
  \caption{Side view}
\end{subfigure}
\caption{The structure and tight-binding hopping parameters for silicene (a), (c)  and bilayer graphene (b), (d). In each case a black vertical arrow shows the direction of the applied electric field.}
\label{structure}
\end{figure}

In this study we use a classification similar to that proposed for single layer graphene QDs.~\cite{Zarenia2011} The structures are classified based on their shape and edge type. As can be seen from Fig.~\ref{fig:Classification}, four types of QD can be distinguished. Depending on their edge geometry, QDs can be classified as the zigzag or armchair QDs that are presented in Fig.~\ref{fig:Classification} (a), (b) and (c), (d), correspondingly. A quantum dot of each of these types can have triangular (TRI) or hexagonal (HEX) shape. The number of atoms in the cluster varies depending on its shape and size. Table~\ref{tab:ConversionTable} summarizes how different size characteristics are connected with the total number of atoms in the single layer structure, $n$, by means of the number of characteristic hexagonal elements and the lattice parameter $a_0$. The choice of a characteristic element for the structure indexing is a matter of convention.  As shown in Fig.~\ref{fig:Classification} by larger and smaller font numbering, one can count hexagons or, equivalently, edge atoms. In the case of a QD with zigzag edges, shown in Fig.~\ref{fig:Classification} (a), (b), edge atoms on a single edge are counted, whereas for QDs with armchair edges, presented in Fig.~\ref{fig:Classification} (c), (d), edge atom pairs are counted.~\cite{Yamamoto2006} The lattice parameter $a_0$ is the distance between the nearest atoms, or their projections onto a horizontal plane as depicted in Fig.~\ref{structure} (d) and (b) for a flat and low-buckled structure, respectively. Apparently, to obtain the total number of atoms, $n_{\text{tot}}$, in a bilayer (multilayer) structure the number of atoms in the Table~\ref{tab:ConversionTable} should be multiplied by the number of layers.



\begin{table*}[htbp]
\caption{ Relations between the number of atoms per layer, $n$, and quantum dot size characteristics: circumscribed circle radius $R$, edge length $L$, and the number of edge atoms $N_z$ (or edge atom pairs $N_{a}$). The parameter $a_0$  is the distance between the nearest atoms in 2D hexagonal lattice or their projection onto a horizontal plane in case of the buckled structure  ($\approx 1.42$ {\AA}  for graphene and $\approx 2.21$~{\AA} for silicene~\cite{Cahangirov2009}). } 
\begin{tabular}{cccccc} \hline \hline
   & \multicolumn{4}{c}{Quantum dot type} \\ 
   & \multicolumn{2}{c}{Zigzag} & \multicolumn{2}{c}{Armchair} \\ 
   & triangular & hexagonal & triangular & hexagonal \\ \hline
   $R$ & $\left(N_z + 1\right) a_0$ & $\sqrt{3} \left(N_z - 1/3\right)\, a_0 $ & $\sqrt{3} N_a a_0$ & $\left(3 N_a - 2\right) a_0 $  \\ 
   $L$ & $\sqrt{3} \left(N_z + 1\right) a_0$~\footnote{\label{footnote1}Ref.~\cite{Zarenia2011}} & $\sqrt{3} \left(N_z - 1/3\right)\, a_0$~\footref{footnote1}  & $3 N_a a_0$  & $\left(3 N_a - 2\right) a_0 $ \\ 
   $n$ & $N_z^2 + 4 N_z +1$~\footnote{\label{footnote2}Ref.~\cite{Yamamoto2006}} & $6 N_z^2$ & $3 N_a \left(N_a + 1\right)$~\footref{footnote2} & $6 \left(3 N_a^2 - 3 N_a + 1\right)$ \\  
   $N_{z,a}$ & $\sqrt{n+3} - 2$ & $\sqrt{\dfrac{n}{6}}$ & $\dfrac{\sqrt{12 n +9} - 3}{6}$ & $\dfrac{\sqrt{2 n - 3} + 3}{6}$ \\ \hline \hline
\end{tabular}
\label{tab:ConversionTable}
\end{table*}

 The electronic properties of presented clusters in a transverse
electric field can be calculated using the tight-binding Hamiltonian\cite{McCann,CLiu2011_2,Ezawa2012},
\begin{equation}
H=\sum_{\left\langle ij\right\rangle }t_{ij}c_{i}^{\dag
}c_{j}+\sum_{i}V_{i}\left( E\right) c_{i}^{\dag }c_{i},  \label{Htb_s}
\end{equation}%
where $c_{i}^{\dag }$ and $c_{i}$ are the electron creation and annihilation
operators, $t_{ij}$ are the inter-site hopping parameters and $V_{i}$ is the
on-site electron potential that depends both on the local atomic environment
and on the applied electric field. The hoping parameters $t_{ij}$ can be
written in terms of the nearest-neighbor (NN) coupling constants $\gamma _{i}$,
as illustrated in Fig.~\ref{structure}.
In the case of silicene we use the simplified version appropriate for the
low-energy states.\cite{Ezawa,CLiu2011_2} According to this approximation
there is only one \textit{in-plane} coupling parameter between sites A and
B, $\gamma _{0}\simeq 1.6~\mathrm{eV}$, that corresponds to the
nearest-neighbor hopping between sites A and B. 

For graphene this parameter is $\gamma _{0}\simeq 3~\mathrm{eV}$.
The on-site potential, $V_{i}(E)$ is different for A and B sites and can be presented as $V_{i}=\xi
_{i}\Delta -\xi _{i}lE$ where $\xi _{i}=\pm 1$ for the B and A type of
atoms, $\Delta \simeq 3.9~\mathrm{meV}$ is the effective buckling-gap
parameter and $lE$ is the field-induced electrostatic interaction, related
to the up/down shift of B and A atoms on $l\simeq 0.22\,\mathring{A}$ with
respect to the average plane.

For the bilayer graphene structure, along with the \textit{in-plane}
coupling $\gamma _{0}\simeq 3.16~\mathrm{eV}$, the \textit{inter-layer}
parameters $\gamma _{1}\simeq 0.38~\mathrm{eV}$, $\gamma _{3}\simeq 0.38~%
\mathrm{eV}$ and $\gamma _{4}\simeq -0.14~\mathrm{eV}$ ( see Fig.~\ref{structure} (b)) should be also taken into account. The field-dependent on-site potential
can be written \cite{ McCann} as $V_{i}=\eta _{i}\Delta -\varsigma _{i}lE$
where $\eta _{i}=0$ for A1 and B2 atoms, $\eta _{i}=1$ for A2 and B1 atoms
and $\varsigma _{i}=\pm 1$ for the atoms located in the upper (A2, B2) and lower
(A1, B1) layers correspondingly (See Fig.~\ref{structure}). The on-site potential due to
the different local atomic environments is taken as $\Delta \simeq 22~%
\mathrm{meV}$ and the inter-layer distance as $2l \simeq 3.5\,\mathring{A}$.

By numerically diagonalizing the Hamiltonian given by Eq.~(\ref{Htb_s}) one finds
the single-electron wave functions $\left\vert \Psi _{i}\right\rangle $
and their corresponding energy levels $\epsilon _{i}$, which can then be used to
evaluate the optical absorption cross section given by the following
expression: 
\begin{equation}
\sigma (\epsilon )\sim \sum_{i,f}S(\epsilon _{i,f})\delta (\epsilon
-\epsilon _{i,f})\,,  \label{sigma}
\end{equation}%
where $S(\epsilon _{i,f})$ is the oscillator strength, and $\delta(\epsilon-\epsilon _{i,f})$ is the Dirac delta function. The oscillator strength characterizing the rate
of transitions between the initial, $\left\vert \Psi _{i}\right\rangle $,
and the final, $\left\vert \Psi _{f}\right\rangle $, states is defined as~%
\cite{Lee2002} 
\begin{equation}
S(\epsilon _{i,f})\sim \epsilon _{i,f}\left\vert \left\langle \Psi
_{f}\left\vert \hat{\vec{r}}\right\vert \Psi _{i}\right\rangle \right\vert
^{2}\,.  \label{eq:OscillatorStrength}
\end{equation}%
In Eq. (\ref{eq:OscillatorStrength}) $\hat{\vec{r}}$ is the position
operator and $\epsilon _{i,f}=\epsilon _{f}-\epsilon _{i}$ is the energy of
a single-electron transition between the states with energies $\epsilon _{i}$
and $\epsilon _{f}$. The summation in Eq.~(\ref{sigma}) is carried out over all
possible transitions between the valence and conduction states.

To mimic thermal level broadening, finite single electron excitation
lifetimes, nanocluster size inhomogeneity, etc., single electron absorption
peaks in Eq. (\ref{sigma}) are broadened by a Gaussian function with
linewidth, $\alpha $, 
\begin{equation}
\sigma (\epsilon )\sim \sum_{i,f}S(\epsilon _{i,f})\exp \left( -\frac{%
(\epsilon -\epsilon _{i,f})^{2}}{\alpha ^{2}}\right) \,,
\end{equation}

As follows from Eqs.(\ref{sigma}) and (\ref{eq:OscillatorStrength}),
calculation of the absorption spectrum is reduced to a calculation of the matrix
elements of the position operator, i.e., $\left\langle \Psi _{f}\left\vert 
\hat{\vec{r}}\right\vert \Psi _{i}\right\rangle $. Within the tight-binding
model in its most general form this physical quantity is given by \cite{Leung1997,Leung1998,Schulz2006} 
\begin{equation}
\begin{split}
\left\langle \Psi _{i}\left\vert \hat{\vec{r}}\right\vert \Psi
_{j}\right\rangle =\sum_{m,\gamma ,\gamma ^{\prime }} C_{i,m,\gamma }^{\ast
}C_{j,m,\gamma ^{\prime }} \vec{r}_{m}\delta _{\gamma ,\gamma ^{\prime }}\\
        +\sum_{m,\gamma ,\gamma ^{\prime }} C_{i,m,\gamma }^{\ast }C_{j,m,\gamma
^{\prime }} \left\langle \phi _{m,\gamma }\left\vert \hat{\vec{r}}-\vec{r}%
_{m}\right\vert \phi _{m,\gamma ^{\prime }}\right\rangle ,
\end{split}
\label{eq:MatrixElement}
\end{equation}%
where $\vec{r}_{m}$ is the position of the $m$-th atom in the QD, $\phi
_{m,\gamma }$ is the atomic orbital $\gamma $ of the {$m$-th} atom, $%
C_{i,m,\gamma }$ are the coefficients of the expansion of the electron
wavefunction in terms of the atomic orbitals. The first sum in Eq.~(\ref{eq:MatrixElement}) is the dipole moment associated with the positions of the atoms of the QD. Due to the orthogonality of the electron wave functions of any two different states the value of this sum does not depend upon the choice of
the origin of the coordinate system. Hence, only the relative atomic positions
with respect to each other contribute to this term and, therefore, it is
usually referred to as the inter-atomic dipole moment. The second sum of Eq. (\ref{eq:MatrixElement}) represents the dipole moment
of transitions between orbitals $\gamma $ and $\gamma ^{\prime }$ located on
the same atomic site and it is usually referred to as intra-atomic dipole
moment. 
The intra-atomic dipole moment restores the result for an isolated
atom in the limit of non-interacting atoms of the QD. In contrast to this,
ZES arise due to the interaction between the atoms. Therefore, the
contribution of intra-atomic dipole moments to the resulting dipole moment
of transitions between low-energy states is assumed to be small. Taking
into account the fact that the low-energy electronic structure of silicene and
bilayer graphene QDs is formed by $\pi$-atomic orbitals, one can reduce Eq. (%
\ref{eq:MatrixElement}) to the following form:
\begin{equation}
\left\langle \Psi _{i}\left\vert \hat{\vec{r}}\right\vert \Psi _{j}\right\rangle
=\sum_{m}C_{i,m}^{\ast }C_{j,m}\vec{r}_{m}\,,
\end{equation}%
where $C_{i,m}$ are the coefficients of expansion of the electron wave
function $\Psi _{i}$ in the basis of the $\pi$-orbitals $\phi _{m}$, 
\begin{equation}
\Psi _{i}=\sum_{m}C_{i,m}\phi _{m}(\vec{r}-\vec{r}_{m})\,.
\label{eq:WaveFunction}
\end{equation}%
The unknown coefficients of Eq.~(\ref{eq:WaveFunction}), $C_{i,m}$, are
the components of eigenvectors of the Hamiltonian given by Eq.~(\ref{Htb_s}).

\begin{figure}[htp]
\centering
\begin{subfigure}{.35\textwidth}
  \centering
  \includegraphics[width=\textwidth]{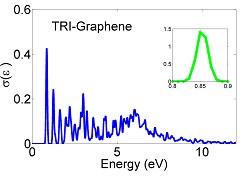}
  \caption{}
\end{subfigure}%

\begin{subfigure}{.35\textwidth}
  \centering
  \hspace{-.5cm}
  \includegraphics[width=\textwidth]{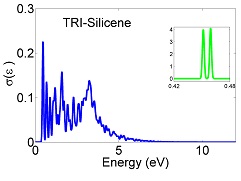}
  \caption{}
\end{subfigure}%

\begin{subfigure}{.35\textwidth}
  \centering
  \includegraphics[width=\textwidth]{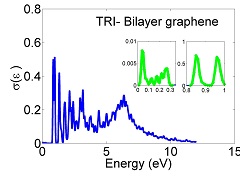}
  \caption{}
\end{subfigure}%
\caption{Optical absorption cross sections of triangular (TRI) quantum dots with zigzag
edges based on (a) graphene, (b) silicene and (c) bilayer graphene. Insets show zoomed in regions of interest. Each cluster has $438$ atoms per layer.}
\label{absorptionGQD}
\end{figure}

\section{\label{Sec:Res} Results and Discussion}

\subsection{Optical absorption of triangular quantum dots}

Optical absorption cross sections per atom, $\sigma (\epsilon )/n_{\text{tot}}$, were obtained in arbitrary units for graphene, silicene, and bilayer
graphene clusters  with $438$ atoms per layer ($L\approx 77$ {\AA} for silicene and $L\approx 49$ {\AA} for single layer and bilayer graphene) by the procedure described in Section~\ref{Sec:Model}. The results are depicted in Fig.~\ref{absorptionGQD}.  The number of ZES in the selected triangular clusters is equal to $18$ for graphene and silicene QDs, and to $36$ in the bilayer
graphene QDs. This number can be expressed in terms of the size parameter $N_z$, specified in Table~\ref{tab:ConversionTable}, as $N_z - 1$ and it should be multiplied by the number of layers for bilayer clusters. In the present calculations and thereafter the optical absorption cross section is a result of transitions from states below to states above the Fermi level.
The linewidth for the main panels in  Fig.~\ref{absorptionGQD} was selected to be equal to $\alpha =45$ meV whereas, for the study of low-energy features (insets in Fig.~\ref{absorptionGQD}), parameter $\alpha $ was selected to be equal to $14$ meV for graphene and bilayer graphene, $4.5$ meV for silicene QDs.

We consider first transitions at zero electric field. The dependence of the optical absorption cross section for
graphene clusters on the transition energy is shown in Fig.~\ref{absorptionGQD} (a). The results are in good agreement with those of Yamamoto
et al.~\cite{Yamamoto2006} Figure~\ref{absorptionGQD} (b) presents the corresponding  $\sigma
(\epsilon )/n_{\text{tot}}$ for silicene QDs. The low-energy zoom at the inset to this
figure reveals the shift of the $0.85$ eV graphene peak towards $0.45$ eV in
silicene as a result of the decrease in the hopping energy. The more important
difference, however, is the splitting of this peak in two peaks. This effect is
caused by the fact that ZES in silicene are no longer localized at $\epsilon
=0$~\cite{Hazem2} and, therefore, the transition energy from the valence states to
the  ZES is different from the transition energy from the
 ZES to the conduction states.

The situation with the low-energy peak changes even more for the case of the
triangular bilayer graphene QD where the ZES are smeared into the narrow
energy band by the inter-layer electron hopping.\cite{Hazem2} This smearing
creates the dispersion of the optical absorption peaks in the region $0-1.0$
eV as shown in the inset of Fig.~\ref{absorptionGQD} (c). These peaks correspond to the
possible transitions from the dispersed ZES and  valance states to the
dispersed ZES in the conduction band. Such a feature exists neither
in graphene nor in silicene single layers where all the ZES are degenerate.

\subsection{Electric field effect and optical absorption}

\subsubsection{Silicene QDs with zigzag edges}
\begin{figure}[htp]
\centering


\begin{subfigure}{.25\textwidth}
  \centering
  \includegraphics[width=\textwidth]{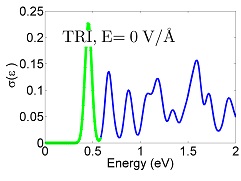}
  \caption{}
\end{subfigure}%
\begin{subfigure}{.25\textwidth}
  \centering
 \centerline{\includegraphics[width=\textwidth]{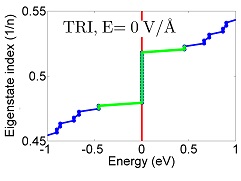}}
  \caption{}
\end{subfigure}

\begin{subfigure}{.25\textwidth}
  \centering
  \includegraphics[width=\textwidth]{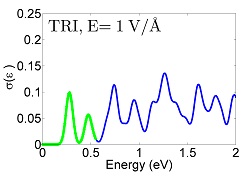}
  \caption{}
\end{subfigure}%
 \begin{subfigure}{.25\textwidth}
  \centering
 \centerline{\includegraphics[width=\textwidth]{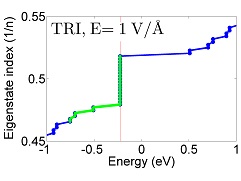}}
  \caption{}
\end{subfigure}%

\begin{subfigure}{.25\textwidth}
  \centering
  \includegraphics[width=\textwidth]{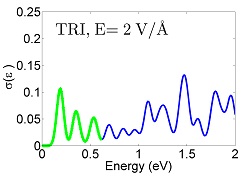}
  \caption{}
\end{subfigure}%
\begin{subfigure}{.25\textwidth}
  \centering
 \centerline{\includegraphics[width=\textwidth]{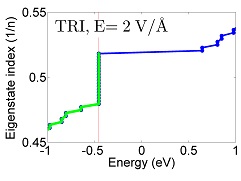}}
  \caption{}
  \label{fig:sub2}
\end{subfigure}

\caption{Optical absorption cross sections (a), (c), (e) and corresponding energy levels (b), (d), (f) for a triangular zigzag silicene QD consisting of  $438$ atoms ($L\approx77$ {\AA}) at different electric fields.}
\label{absorptionTSQD}
\end{figure}

Figure~\ref{absorptionTSQD} illustrates the effect of the electric field, $E$,
on the optical absorption (a), (c), (e) and on the energy levels spectrum (b), (d), (f) of 
triangular silicene QDs. As can be seen from Fig.~\ref{absorptionTSQD}(a), there is only one absorption peak below the energy $\epsilon =0.5$ eV when $E=0$. This peak includes two types of transitions:
from the highest occupied energy level (HOEL) to the
ZES and from the ZES to the lowest unoccupied energy level (LUEL). In graphene these two types of transitions have the same transition energy but in silicene they are not identical and the energy
difference between them, which is zero at $E=0$, can be tuned by the
electric field. 

With increasing electric field two remarkable effects occur. Firstly, the two
indicated transitions become non-identical, which results in splitting the
corresponding peak in two peaks. The first peak lies below $\epsilon \simeq 0.5$ eV at $E=1$ V/{\AA}, the first green peak in Fig.~\ref{absorptionTSQD}(c), while the second peak is
positioned at the higher energy. Secondly, ZES become closer to the valence
band states. Therefore transitions from some of the valence band states to
ZES appear at the energies below $\epsilon \simeq 0.5$ eV. These
transitions are represented by the second green peak in Fig.~\ref{absorptionTSQD}(c). In the higher field, $E=2$ V/{\AA }, the energy difference between ZES and valence band states becomes even smaller which results in the appearance of the third green peak in Fig.~\ref{absorptionTSQD}(c). For a negative electric field the behaviour is similar but one should note that the absorption peaks at $\epsilon \simeq 0.5$ eV are now a result of transitions from ZES to the conduction band states.
\begin{figure}[htp]
\centering
\begin{subfigure}{.25\textwidth}
  \centering
  \includegraphics[width=\textwidth]{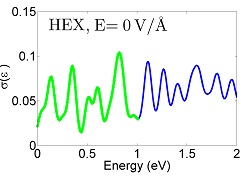}
  \caption{}
\end{subfigure}%
\begin{subfigure}{.25\textwidth}
  \centering
 \centerline{\includegraphics[width=\textwidth]{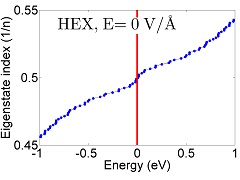}}
  \caption{}
\end{subfigure}
\begin{subfigure}{.25\textwidth}
  \centering
  \includegraphics[width=\textwidth]{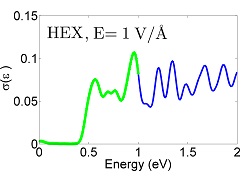}
  \caption{}
\end{subfigure}%
\begin{subfigure}{.25\textwidth}
  \centering
 \centerline{\includegraphics[width=\textwidth]{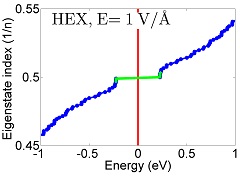}}
  \caption{}
\end{subfigure}
\begin{subfigure}{.25\textwidth}
  \centering
  \includegraphics[width=\textwidth]{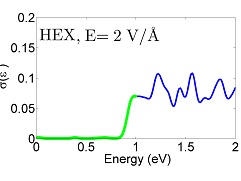}
  \caption{}
\end{subfigure}%
\begin{subfigure}{.25\textwidth}
  \centering
 \centerline{\includegraphics[width=\textwidth]{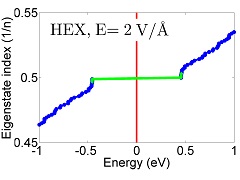}}
  \caption{}
\end{subfigure}
\caption{The same as Fig.~\ref{absorptionTSQD}, but for a hexagonal (HEX) silicene QD containing $864$ atoms ($L\approx 45$ {\AA}) at different electric fields. }
\label{absorptionHSQD}
\end{figure}

Hexagonal silicene QDs have no ZES. Therefore, the effect of the electric
field is just in the opening of a tunable energy gap.\cite{Hazem2} This is
clearly seen in the optical absorption spectra as a shift of the edge of the
absorption in Fig.~\ref{absorptionHSQD}(a), (c), (e). Without the electric field the
absorption peaks are distributed  almost uniformly in the region of $0 - 1$
eV. However, application of the field results in their shifting to the higher
energies and in the emergence of an energy region with zero absorption.
Thus, one can distinguish two regions with zero and non-zero absorption. Note
also that the intensity of the peak near the absorption edge depends on the
field. The increase of the electric field from $1$ to $2$~V/{\AA} results in a gentle decrease of the peak.
\begin{figure}[htbp]
\centering
\begin{subfigure}{.25\textwidth}
  \centering
  \includegraphics[width=\textwidth]{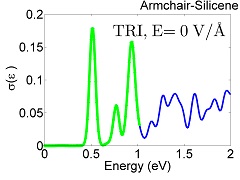}
  \caption{}
\end{subfigure}%
\begin{subfigure}{.25\textwidth}
  \centering
  \centerline{\includegraphics[width=\textwidth]{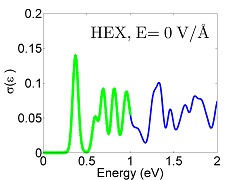}}
  \caption{}
\end{subfigure}
\begin{subfigure}{.25\textwidth}
  \centering
  \includegraphics[width=\textwidth]{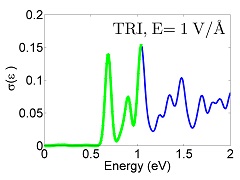}
  \caption{}
\end{subfigure}%
\begin{subfigure}{.25\textwidth}
  \centering
 \centerline{\includegraphics[width=\textwidth]{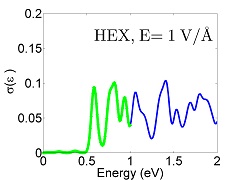}}
  \caption{}
\end{subfigure}
\begin{subfigure}{.25\textwidth}
  \centering
  \includegraphics[width=\textwidth]{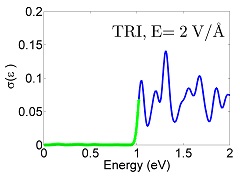}
  \caption{}
\end{subfigure}%
\begin{subfigure}{.25\textwidth}
  \centering
 \centerline{\includegraphics[width=\textwidth]{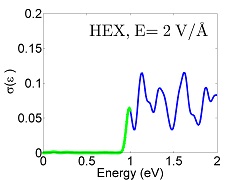}}
  \caption{}
\end{subfigure}
\caption{Optical absorption spectra for armchair silicene QDs of  triangular (a), (c), (e) and hexagonal (b), (d), (f) shapes, consisting of $468$ and $762$ atoms and having edge length $L \approx 80$~{\AA} and $L \approx 42$~{\AA}, respectively.}
\label{absorptionASQD}
\end{figure}

\subsubsection{Silicene QDs with armchair edges}
In order to present the effect of edge termination on the electronic and optical properties of the silicene we  extend our calculations to account for silicene flakes with armchair edges. The optical absorption cross sections of triangular and hexagonal silicene QDs with armchair edges are shown in Fig. \ref{absorptionASQD}. The total numbers of atoms are: $n_{\text{tot}}=468$ and $n_{\text{tot}}=762$ atoms ($L \approx 80$ {\AA} and $L \approx 42$ {\AA} ) for triangular and hexagonal flakes, respectively. At zero electric field, see Fig. \ref{absorptionASQD}(a), the absorption spectrum for triangular armchair looks similar to the spectrum of triangular zigzag, see Fig. \ref{absorptionTSQD}(a), with one absorption peak around $\epsilon=0.5$ eV. However, applying an electric field to triangular armchair clusters does not shift the absorption edge to the lower energy as in zigzag clusters. It is clearly seen in Fig.~\ref{absorptionASQD}(a), (c), and (e) that the absorption edge blue shifts with the application of an electric field. The  reason for such a behaviour is the absence of ZES in armchair silicene flakes. The shifting of the ZES in zigzag flakes closer to the conduction band or to the valance band decreases the energy gap. 
Unlike zigzag hexagonal QDs, armchair haexagonal clusters at $E=0$ have a significant energy gap $\simeq 0.3$ eV. As indicated by Fig. \ref{absorptionASQD}(b), (d), and (f), in an electric field this gap increases similar to that opened by the field in the zigzag clusters.

\subsubsection{Bilayer graphene QDs with zigzag edges}
\begin{figure}[htbp]
\centering
\begin{subfigure}{.25\textwidth}
  \centering
  \includegraphics[width=\textwidth]{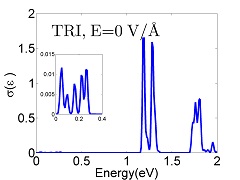}
  \caption{}
\end{subfigure}%
\begin{subfigure}{.25\textwidth}
  \centering
  \centerline{\includegraphics[width=\textwidth]{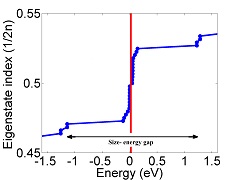}}
  \caption{}
\end{subfigure}
\begin{subfigure}{.25\textwidth}
  \centering
  \includegraphics[width=\textwidth]{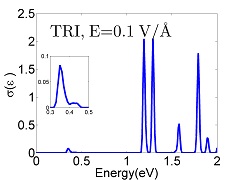}
  \caption{}
\end{subfigure}%
\begin{subfigure}{.25\textwidth}
  \centering
 \centerline{\includegraphics[width=\textwidth]{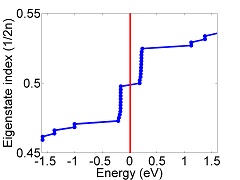}}
  \caption{}
\end{subfigure}
\begin{subfigure}{.25\textwidth}
  \centering
  \includegraphics[width=\textwidth]{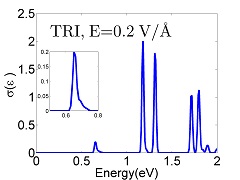}
  \caption{}
\end{subfigure}%
\begin{subfigure}{.25\textwidth}
  \centering
 \centerline{\includegraphics[width=\textwidth]{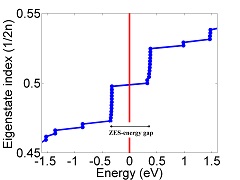}}
  \caption{}
\end{subfigure}
\caption{Optical absorption spectra  (a), (c), (e) and energy levels
 (b), (d), (f) of a triangular bilayer graphene QD  made of $222$ atoms per layer ($L \approx 34$~{\AA}) at different electric fields. The insets show zoomed in absorption peaks below $0.8$ eV.}
\label{absorptionBGQD}
\end{figure}

In this section the clusters of triangular and hexagonal bilayer graphene QDs  
with number of atoms per layer $n=222$ and $n=216$, respectively, are considered.
We study the optical properties of the triangular bilayer graphene QDs,
whose energy levels are shown in Fig.~\ref{absorptionBGQD}(b). 
At zero field, ZES can be divided into two groups. The first group represents ZES located below the Fermi level, left side of the red line in Fig.~\ref{absorptionBGQD}(b), at $\epsilon \simeq -0.1$ eV. The second group represents ZES located above the Fermi level at $\epsilon \simeq 0.1$ eV. Then we study the optical absorption peaks resulting from the transitions between these two groups under the effect of electric field. In general, the smearing of ZES and the application of an electric field affects all optical transitions from and to ZES but we focus here only on the transitions between the two previously discussed groups of ZES. These transitions can be seen in Fig.~\ref{absorptionBGQD}(a) in the energy range from $0$ to $0.3$ eV. Thus, one can identify one group  of optical transitions within the ZES. The inset of Fig.~\ref{absorptionBGQD} (a) at $E=0$ V/{\AA} shows a series of absorption peaks in the energy range from $0$ to $0.3$ eV. These small intensity peaks represent the group of transitions mentioned. The application of the electric field increases the tiny energy gap in the middle of the ZES band and gathers the ZES groups into the narrower energy range. 
 \begin{figure}[htp]
\centering
\begin{subfigure}{.25\textwidth}
  \centering
  \includegraphics[width=\textwidth]{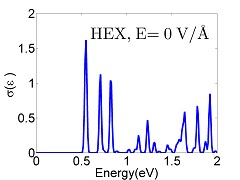}
  \caption{}
\end{subfigure}%
\begin{subfigure}{.25\textwidth}
  \centering
  \centerline{\includegraphics[width=\textwidth]{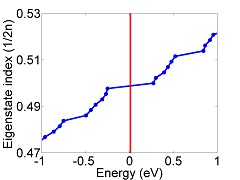}}
  \caption{}
\end{subfigure}
\begin{subfigure}{.25\textwidth}
  \centering
  \includegraphics[width=\textwidth]{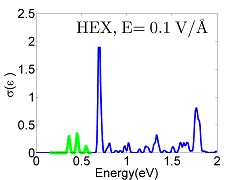}
  \caption{}
\end{subfigure}%
\begin{subfigure}{.25\textwidth}
  \centering
 \centerline{\includegraphics[width=\textwidth]{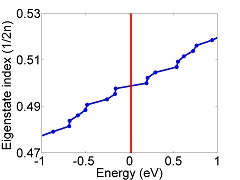}}
  \caption{}
\end{subfigure}
\begin{subfigure}{.25\textwidth}
  \centering
  \includegraphics[width=\textwidth]{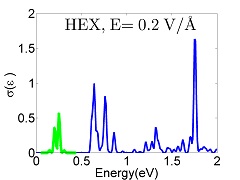}
  \caption{}
\end{subfigure}%
\begin{subfigure}{.25\textwidth}
  \centering
 \centerline{\includegraphics[width=\textwidth]{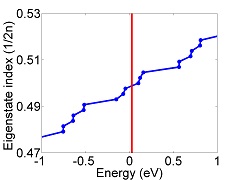}}
  \caption{}
\end{subfigure}
\caption{The same as Fig.~\ref{absorptionBGQD}, but for a hexagonal bilayer graphene QD containing $216$ atoms per layer ($L \approx 14 $~{\AA}) at different electric fields.}
\label{absorptionBGQD_HEX}
\end{figure}
 This leads to the up-shift of the  set of the low-energy absorption peaks to $\epsilon \simeq 
 0.3-0.45$ eV (for $E=0.1$ V/{\AA} ) with the gathering of the small intensity peaks and increase in the peak intensity as shown in the insets of Fig.~\ref{absorptionBGQD}(c) at $E=0.1$ V/{\AA}. Increasing the electric field to $E= 0.2$ V/{\AA} results in a further increase in the energy gap which in turn increases the intensity and the up-shift of the absorption peak to $\epsilon \simeq 0.6-0.7$ eV as can be seen in the inset of Fig.~\ref{absorptionBGQD}(e).
 
 The optical absorption cross section and energy levels for a hexagonal bilayer graphene QD at different values of the electric field are shown in Fig.~\ref{absorptionBGQD_HEX}(a), (c), (e) and Fig.~\ref{absorptionBGQD_HEX}(b), (d), (f), respectively. In deep contrast to triangular bilayer graphene QDs, the energy gap in hexagonal bilayer graphene between the HOEL and LUEL, which is presented in Fig.~\ref{absorptionBGQD_HEX}(b) at E=0 V/{\AA}, decreases with increasing electric field. This feature causes the shifting of some of the absorption peaks marked by green in Fig.~\ref{absorptionBGQD_HEX}(c), (e) to a lower energy and into the energy gap region. At $E=0.1$ V/{\AA}  two low-energy absorption peaks appear in the energy gap region at $\epsilon \simeq 0.3-0.45$ eV.  Increasing $E$ to $0.2$ V/{\AA} leads to further shifting of the absorption peaks to lower energy $\epsilon \simeq0.15-0.3$ eV. Therefore, in small clusters we have blue/red shift for the low-energy absorption peaks for triangular/hexagonal bilayer graphene QDs. 
 \begin{figure}[htp]
\centering
\begin{subfigure}{.25\textwidth}
  \centering
  \includegraphics[width=\textwidth]{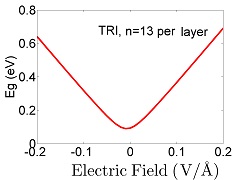}
  \caption{}
\end{subfigure}%
\begin{subfigure}{.25\textwidth}
  \centering
  \centerline{\includegraphics[width=\textwidth]{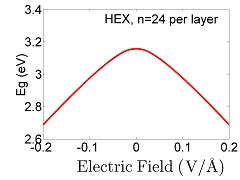}}
  \caption{}
\end{subfigure}
\begin{subfigure}{.25\textwidth}
  \centering
  \includegraphics[width=\textwidth]{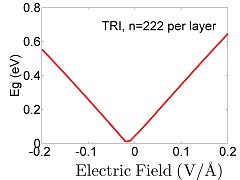}
  \caption{}
\end{subfigure}%
\begin{subfigure}{.25\textwidth}
  \centering
 \centerline{\includegraphics[width=\textwidth]{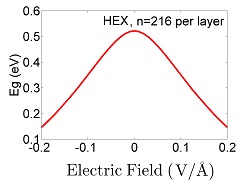}}
  \caption{}
\end{subfigure}
\begin{subfigure}{.25\textwidth}
  \centering
  \includegraphics[width=\textwidth]{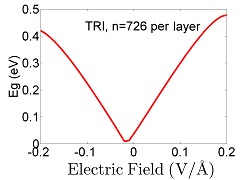}
  \caption{}
\end{subfigure}%
\begin{subfigure}{.25\textwidth}
  \centering
 \centerline{\includegraphics[width=\textwidth]{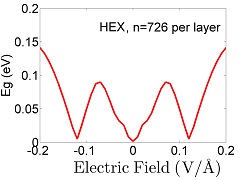}}
 \caption{}
\end{subfigure}
\caption{Energy gap dependence on the applied electric field in  triangular (a) $L \approx 7.4$~{\AA}, (c) $L \approx 34$~{\AA}, (e) $L \approx 64$~{\AA} and hexagonal (b) $L \approx 4.1$~{\AA}, (d) $L \approx 14$~{\AA}, (f) $L \approx 26$~{\AA} bilayer graphene QDs.}
\label{EG_S_BG}
\end{figure}

In order to test this feature for different sizes of triangular and hexagonal bilayer graphene QDs we 
plotted Fig.~\ref{EG_S_BG}, which illustrates the variation in energy gap upon application of electric field in clusters of different sizes. It can be seen from  Fig.~\ref{EG_S_BG}(b), (d), (f) that the energy gap in 
hexagonal clusters decreases with the application of electric field for small clusters and starts to 
increase with the field for a cluster size, where the total number of atoms is $n=726$ per layer. The energy gap for triangular clusters increases with the field for all the selected sizes as seen from Fig.~\ref{EG_S_BG}(a), (c), (e).

We can differentiate two energy gaps. Fhe first one is the size dependent (size-energy gap) shown in Fig.~\ref{absorptionBGQD}(b) and the second is the stacking induced energy gap which occurs between ZES (ZES-energy gap) as indicated in Fig.~\ref{absorptionBGQD}(f). Due to the coupling parameters $\gamma_{4}$ and the on-site potential $\Delta$ in bilayer graphene, the ZES states split into two groups  giving rise to the ZES-energy gap which is directly proportional to the applied electric field. If we consider the size-energy gap, shown in Fig.~\ref{absorptionBGQD}(b), we can easily see from Fig.~\ref{absorptionBGQD}(b), (d), and (f) that this energy gap decreases by increasing the electric field. Therefore the decreased energy gap as a function of electric field in HEX bilayer small QDs (Fig.~\ref{EG_S_BG}(b), (d)) can be attributed to the absence of  ZES in small clusters of HEX graphene bilayer QDs. Thus, there is only the size-energy gap which decreases in a similar manner to TRI bilayer graphene QDs. As shown in Fig. \ref{EG_S_BG}(f) increasing the size of the HEX cluster leads to oscillation of the energy gap as a function of the electric field. In fact, the oscillatory behaviour of the energy gap with electric field occurs not only for cluster with $n=726$ and higher but also for smaller clusters at higher values of the electric field. 
It has been reported  recently that hexagonal bilayer graphene QDs with zigzag edges in the presence of an electric field  exhibit unusual edge states inside the energy gap; these states oscillates as the applied electric field increases.~\cite{Costa2015} Therefore, according to the previous results\cite{Costa2015} and our results, we conclude that the energy gap oscillation is a result of the oscillation of the edge states inside the gap due to increasing the electric field.

\subsubsection{Bilayer graphene QDs with armchair edges}
\begin{figure}[htbp]
\centering
\begin{subfigure}{.25\textwidth}
  \centering
  \includegraphics[width=\textwidth]{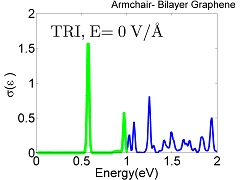}
  \caption{}
\end{subfigure}%
\begin{subfigure}{.25\textwidth}
  \centering
  \centerline{\includegraphics[width=\textwidth]{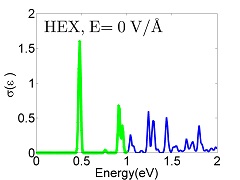}}
  \caption{}
\end{subfigure}
\begin{subfigure}{.25\textwidth}
  \centering
  \includegraphics[width=\textwidth]{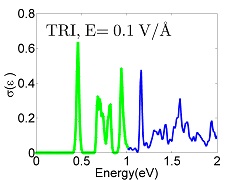}
  \caption{}
\end{subfigure}%
\begin{subfigure}{.25\textwidth}
  \centering
 \centerline{\includegraphics[width=\textwidth]{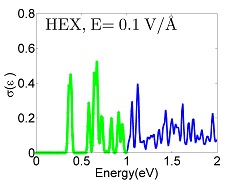}}
  \caption{}
\end{subfigure}
\begin{subfigure}{.25\textwidth}
  \centering
  \includegraphics[width=\textwidth]{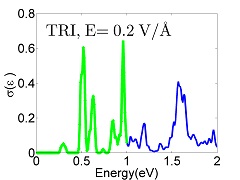}
  \caption{}
\end{subfigure}%
\begin{subfigure}{.25\textwidth}
  \centering
 \centerline{\includegraphics[width=\textwidth]{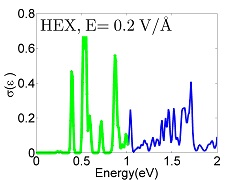}}
  \caption{}
\end{subfigure}
\caption{Optical absorption spectra for armchair bilayer graphene QDs of  triangular (a), (c), (e) and hexagonal (b), (d), (f) shapes with $720$ ($L \approx 64$~{\AA})  and $762$ ($L \approx 27$~{\AA}) atoms per layer, respectively.}
\label{absorption_BGSQD}
\end{figure}

As discussed above for zigzag bilayer clusters, the  increase (decrease) in the absorption gap can be obtained through applying an electric field to triangular (hexagonal) bilayer graphene QDs. Bilayer graphene QDs with armchair termination do not support edge states, thus it is expected that armchair triangular and hexagonal bilayer QDs will follow a similar trend to that obtained in hexagonal zigzag QDs. Figure~\ref{absorption_BGSQD} illustrates the optical absorption cross section of triangular [Fig.~\ref{absorption_BGSQD} (a), (c), (e)] and hexagonal [Fig.~\ref{absorption_BGSQD} (b), (d), (f)] bilayer graphene QDs at different values of electric field. We notice that for triangular and hexagonal  clusters at  E=0 V/{\AA} there are two absorption peaks in the energy range from $0$ to $1$~eV. Application of an electric field leads to increase in the number of absorption peaks in this energy region and a small shift to lower energy. 

\section{\label{Sec:Conc} Conclusions}

The optical absorption spectra of silicene and bilayer graphene QDs have been investigated for triangular and hexagonal clusters and compared to the corresponding clusters of monolayer graphene. Silicene QDs in a zero electric field show two optical transition peaks from HOEL in the valence band to ZES and from ZES to the LUEL in the conduction band. In contrast to this in graphene QDs these two optical transitions are identical and produce one absorption peak because HOEL and LUEL are symmetric with respect to the ZES. In general, doubling the number of transition peaks occurs not only for transitions from HOEL to ZES and from ZES to LUEL but also for all transitions from all valence band energy states to ZES and from ZES to all conduction band states. Without the electric field, triangular bilayer graphene QDs exhibit optical transitions between the ZES themselves due to the smearing of their ZES. These transitions do not exist in graphene or silicene QDs. 

The introduction of an electric field into silicene triangular QDs displaces the ZES in the energy gap such that they are closer to the conduction states in the case of a positive electric field and push them further away for a negative field. This displacement increases with increasing electric field, thereby increasing the number of absorption peaks in the low-energy region of the optical absorption spectrum. By contrast,  hexagonal silicene QDs show a reduction in the number of optical absorption peaks in the low-energy region with increasing electric field in either direction. 
In triangular bilayer graphene QDs the small energy gap between the ZES increases with increasing electric field. As a result of these field-dependent energy gaps, the edge of absorption due to transitions between ZES undergoes blue shift in response to the applied field. For small clusters of hexagonal bilayer graphene, the edge of absorption has a red shift with increasing electric field. 

Armchair flakes of silicene and bilayer graphene exhibit a significant dependence of their optical properties with electric field. The blue (red) shift of the absorption edge takes place for silicene (bilayer graphene) flakes for both hexagonal and triangular shapes. The absence of ZES in armchair flakes removes the ability to switch the trend of energy gap dependence on the electric field by changing the shape between triangular and hexagonal. Therefore ZES provide a privilege in silicene and bilayer graphene QDs with zigzag edges over those with armchair edges in controlling the electronic and optical properties using different shapes.

The results of the present study should be supplemented in the future by more sophisticated models which 
take into account electron-electron interaction. For instance, the low-energy absorption of the bilayer clusters in conjunction with the magnetic phase transition~\cite{Guclu2011}, depending on the value of the applied electric field,  is worth special attention since in this case electron-electron interaction may results in emergence of additional low-energy transitions. However, we expect that this will not change much the revealed general trends. For instance, electron-electron interaction should not drastically affect such reported features as the highly tunable absorption peak centered at about $0.5$ eV for zigzag silicene QDs of triangular shape. The on-site Coulomb repulsion, omitted in the present consideration, should increase the splitting of the peaks caused by the spin-orbit term in silicene and parameters $\gamma_4$ and $\Delta$ in the bilayer graphene QDs for transitions between ZES and the valence (conduction) band states.

Thus, we have shown that optical spectroscopy in an applied electric field provides a powerful tool for determining the shape and size of the small clusters of silicene and bilayer graphene. In addition, our results provide the basis for using small silicene and bilayer graphene clusters as active elements of mid-infrared optoelectronic devices tunable by an external electric field.



\begin{acknowledgments}
This work was supported by the EU FP7 ITN NOTEDEV (Grant No. FP7-607521); EU H2020 RISE project CoExAN (Grant No. H2020-644076); FP7 IRSES projects CANTOR (Grant No. FP7-612285), QOCaN (Grant No FP7-316432), InterNoM (Grant No FP7-612624); Graphene Flagship (Grant No. 604391) and the Egyptian mission sector. The authors are thankful to Dr. C. A. Downing from IPCMS for helpful discussions and hospitality in Strasbourg and R. Keens for a careful reading of the manuscript.
\end{acknowledgments}

\end{document}